\documentstyle[12pt]{article}

\large
\newcommand{\be}{\begin{equation}}
\newcommand{\ee}{\end{equation}}
\newcommand{\bea}{\begin{eqnarray}}
\newcommand{\eea}{\end{eqnarray}}
\newcommand{\bd}{\begin{displaymath}}
\newcommand{\ed}{\end{displaymath}}
\newcommand{\bi}{\begin{itemize}}
\newcommand{\ei}{\end{itemize}}
\newcommand{\bc}{\begin{center}}
\newcommand{\ec}{\end{center}}
\newcommand{\bfl}{\begin{flushleft}}
\newcommand{\efl}{\end{flushleft}}
\newcommand{\bfr}{\begin{flushright}}
\newcommand{\efr}{\end{flushright}}
\newcommand{\f}{\frac}

\def\6{\partial}  
 \def\d{\delta}  \def\e{\epsilon}

  \def\x{\xi} \def\p{\pi}
 \def\ss{\sigma}

 \def\G{\Gamma} \def\D{\Delta}
\def\Th{\Theta}

\def\={\!\!\!&=&\!\!\!}
\def\+{\!\!\!&&\!\!\!+~}
\def\-{\!\!\!&&\!\!\!-~}

\begin{document}
\title{Flow-Equation method for a superconductor with magnetic correlations }
\author{C.P.Moca\\Dept. of Physics, Univ. of Oradea\\ M.Crisan and I.Tifrea\\
Dept. of Theoretical Physics, "Babes-Bolyai" University, Cluj}
\maketitle
\abstract{ The flow equation method has been used to calculate
the energy of single impurity in a superconductor for the Anderson model
with $U\ne 0$.We showed that the energy of the impurity depends only of the
$\Delta_R^2 $ (renormalized order parameter) which depends of the renormalized
Hubbard repulsion $U^R$. For a strong Hubbard repulsion $U^R=U$ and
$\D^R=\D^I$ the effect of the $s-d$ interactions are nonrelevant,a
result which is expected for this model

}
\maketitle
\newpage
\section{Introduction}

The flow equation method given by Wegner \cite{1} has been successfully applied
for the many-body problem by Kehrein and Mielke \cite{2}, for the Anderson
Hamiltonian. In a previous paper the present authors \cite{4} showed that this
method can be used to calculate the energy of a superconductor containing
magnetic impurities describe by the Anderson Hamiltonian. with $U=0$.
We studied (See ref.\cite{4})the influence of the density of states on the single
impurity energy for the case of a van-Hove density of states. In this case
the energy is reduced by the superconducting state and corrections
depends on $\D^2$. In this paper we consider a superconductor with a
constant density of states but for the impurity we take the Hubbard repulsion
$U\ne 0 $.
\section{Model}

We consider a superconductor containing magnetic impurities describe by
the model Hamiltonian.:
\be
H=H_{BCS}+H_A
\ee
where $H_{BCS}$ is
\be
H_{BCS}=\sum_{k,\ss}\e_k c_{k,\ss}^+c_{k,\ss}+
\sum_k \D(c_{k,\uparrow}^+c_{-k,\downarrow}^+ + c_{k,\uparrow}c_{-k,\downarrow})
\ee
$\D$ being the order parameter  and $H_A$ the Anderson Hamiltonian.
describing the impurity in a metal,as:
\be
H_A=\sum_d \e_d d_{\ss}^+ d_{\ss}+
\sum_{k,\ss}V_{k,d}(c_{k,\ss}^+d_{\ss}+d_{\ss}^+ c_{k,\ss})+
Ud_{\uparrow}^+d_{\downarrow}^+d_{\downarrow}d_{\uparrow}
\ee
In equation (3) the first term is the energy of the impurities,the second
term is the interaction between the itinerant-electron and impurity
and the last term in the Hubbard repulsion between the d-electrons of the
impurity.
In order to make the problem analytically tractable we consider the case
of a one impurity problem.

This problem has been treated using the flow-equation method by Crisan et.al.
\cite{4}for $U=0 $ and a van-Hove density of states. In the next section we
consider the case $U\ne 0$ which is a more realistic case.

\section{The Flow Equations}

The flow equations method,which is in fact a  renormalization procedure
applied in the Hamiltonian formalism has been applied in the solid-state
theory by the Wegner\cite{1}, Kehrein and Mielke \cite{3} and has the main
point the diagonalization of the Hamiltonian which describes the system by a
continuous unitary transformation $\eta(l)$ which lead to a Hamiltonian.
$H(l)$ with the parameters functions of the flow parameter $l$.
This transformation satisfies:
\be
\f{dH(l)}{dl}=[\eta(l),H(l)]
\ee
where $\eta(l)$ can be calculated from:
\be
\eta(l)=[H_0(l),H_{int}(l)]
\ee

In order to solve the differential equations using the Hamiltonian. (1)
we introduce, following  \cite{2} the initial values:
\bea
\e_k^I(l) &  = &  \e_k(l=0 )\nonumber \\
\e_d^I (l)&   =   &  \e_d(l=0)\nonumber \\
 V^I(l)     &  = &  V(l=0)        \\
 U^I(l)    &   =   &  U(l=0)   \nonumber
\eea
and the renormalized value for $l\rightarrow \infty $
\bea
\e_k^R(l) &  = &  \e_k(\infty)\nonumber \\
\e_d^R(l)&   =   &  \e_d(\infty)\nonumber \\
 V^R(l)     &  = &  V(\infty)        \\
 U^R(l)     &   =   &  U(\infty)   \nonumber
\eea
Using the Eq. (5) and the general method (see Ref.\cite{1}) we
calculate $\eta(l) $ as:
\be
\eta(l)=\eta^{(0)}(l)+\eta^{(1)}(l)+\eta^{(2)}(l)+\eta^{(3)}(l)+\eta^{(4)}(l)
\ee
and obtain:
\bea
\eta^{(0)}  & =  & \sum_{k,\ss}\eta_k(c_{k,\ss}^+d_{\ss}-d_{\ss}^+c_{k,\ss})
                 \nonumber\\
            & +   & \sum_{k,\ss}\x_{k}(d_{-\ss}^+c_{k,\ss}^+ +c_{k,\ss}^+ d_{-\ss})
\eea
where:
\be
\eta_k =(\e_k-\e_d)V_{k,d}
\ee
and:
\be
\x_{k,\ss}=-\D V_{-k}\ss
\ee
In Eq.(11) $\ss=\uparrow,\downarrow $ correspond to $\ss=\pm 1$ in the right
side. The higher order contributions will be given as functions of expressions
given by Eqs.(10) and (11) and by:
\be
\Th_{k,\ss}=\eta_{-k}\D\ss +\x_{k,\ss}\e_d
\ee
as:
\be
\eta_{k,\ss}^{(1)}=(\e_k V_k-\e_d V_k-\D \Th_{-k,\ss}\ss)
\ee

\be
\eta_{k,k_1,\ss}^{(1)} =\e_k (\eta_{k,\ss} V_{k_1}+\eta_{k_1,\ss} V_{k})-
                       \D (\x_{-k,\ss} V_{k_1}-\x_{k_1,\ss} V_{-k})\ss
\ee
\be
\eta_{k,\ss}^{(2)}  =  (-\D V_{-k}\ss -\e_k\Th_{k,\ss}-\e_d \Th_{k,\ss} )
\ee
\be
\eta_{k,k_1,\ss}^{(1)}= [\D (\eta_{k_1 \ss}V_k +\eta_{k \ss}V_{k_1})\ss
                             +\e_k(\x_{k1, \ss}V_{-k}+\x_{-k, \ss}V_{k_1})     ]
\ee
\be
\eta_{k,\ss}^{(3)}=(\e_k \eta_{k,\ss}U-\e_d\eta_{k,\ss}U+\D U\x_{-k,\ss}\ss)
\ee
\be
\eta_{k,\ss}^{(4)}=(\D\eta_{-k,\ss}U\ss -\e_k\x_{k}U -\e_d\x_{k}U )
\ee
If we take the spin orientation as $\ss =1$ (the non-magnetic states) the flow equations are:
\bea
\f{d\e_d}{dl}  & = & -2\sum_k\eta_k^{(1)}V_k+2\sum_k\eta_k^{(3) }V_k n_k\nonumber\\
\f{d V_k}{dl}  & = & \eta_k^{(1)} [\e_k-\e_d +\f{U\D^2}{[U(1-n_k)+\e_d+\e_k
                      ][\e_d-\e_k+U]-\D^2  } ]\nonumber\\
\f{d U}{dl}    & = & -4\sum_k\eta_k^{(3) }V_k\\
\f{d \D}{dl}   & = & \f{1}{N(0)} \sum_k\eta_k^{(2)}V_{-k}\nonumber
\eea
where $n_k$ is the Fermi function.
\section{Solutions of the flow equations}

Using the spectral function
\be
J(\e,l )=\sum_kV_k^2\d(\e-\e_k(l) )
\ee
and the factorization $\eta_k^{(1) }(l)=V_k f(\e_k,l) $  the Eqs. (17)  becomes
as follows:
\be
\f{d\e_d}{dl} -\int d\e\f{\6 J(\e,l) }{\6 l}
      \f{ [\e_d-\e+U_1][\e_d+\e+U_2]-\D^2}
        {[\e_d-\e+U][\e_d+\e+U_2][\e_d-\e]-\D^2(\e_d-\e-U) }
\ee
where:
\bea
U_1  & = & U(1+n(\e) ) \nonumber\\
U_2  & = & U(1-n(\e) )
\eea
The equation for $U$ becomes:
\be
\f{d U}{dl}=2\int d\e \f{\6 J(\e,l) }{\6 l }
\f{U[U_2+\e_d+\e] }{[\e_d-\e+U][\e_d-\e][U_2+\e_d+\e]-\D^2[\e_d-\e-U] }
\ee
These equations contain $\D^2$ so we have to transform the equation for $\D$
as:
\be
\f{d \D^2}{dl}=\f{1}{N(0)}\int d\e \f{\6 J(\e,l) }{\6 l }
\f{2U\D^2n(\e) }{[\e_d-\e+U][\e_d-\e][U_2+\e_d+\e]-\D^2[\e_d-\e-U] }
\ee
In the Eqs. (22),(23),(24)we have $\f{\6 J(\e,l) }{\6 l} $ which is
obtained from Eqs. (18) as:
\be
\f{\6 J(\e,l) }{\6 l }=2J(\e,l) f(\e,l)[\e_d-\e +
                      \f{U\D^2}{[\e_d-\e+U][U_2+\e_d+\e]-\D^2}]
\ee
and because we have this relation, a supplementary equation for $V_k(l) $ gives
no more information about the system. The solutions of these equations will
be obtained at $T=0$ $(n(\e)=1-\Th(\e)) $ and if we take a concrete
form for the $J(\e,l=0) $ as
\be
J(\e,l=0)=\f{2V^2}{\p D}=\f{\Gamma}{\p}
\ee
$ \Gamma =\f{2V^2}{\pi} $ where $D$ is the bandwidth in the limit
$U>> D>> \e_d^R$ we obtained from Eqs.(19) using conditions (6),(7):
\bea
\e_d^I    &  = & \e_d^R-\f{\G}{2\p}\f{\D_R^2}{\e_d^{R 2} }
               [1-\f{\e_d^R}{D}+arctanh\f{D}{\e_d^R}+ln\f{D}{\e_d^R}  ]\nonumber\\
\D_I^{ 2} & =  & \D_R^2(1+I_1)\\
U^I       &  = & U^R +\f{2\G}{\p}[ln\f{\e_d+D}{\e_d-D}+ln\f{\e_d+U^R-D}{\e_d^R+U^R+D}  ]
          \nonumber
\eea
where:
\bea
I_1 & = & -\f{1}{2U^R(\e_d^R+U^R)}[2ln\f{\e_d^R+U^R+D}{\e_d^R+D}
          +ln\f{\e_d^{R 2} }{\e_d^{R 2}-D^2 } ] \nonumber\\
    &   &+\f{2(\e_d^R+U^R) }{2U^R(2\e_d^R+U^R ) }arcth \f{D}{\e_d^R}
\eea
\section{Conclusions}

Using the flow equations method we showed that for a BCS superconductor 
with magnetic correlations describe by the Anderson Hamiltonian with $U\ne 0$
we calculated the energy of impurity $\e_d$ the order parameter $\D$ and th e
energy $U$. For a large $D$ we get 
\be
\D_R=\D_I \hspace{3cm}U^R=U^I
\ee 
and th energy of the impurity presents a small variation as function of $D$.

\end{document}